\documentclass[prd,preprintnumbers,amsmath,amssymb]{revtex4}

\usepackage{amsmath}
\newcommand{\bea}{\begin{eqnarray}}
\newcommand{\eea}{\end{eqnarray}}
\newcommand{\be}{\begin{eqnarray}}
\newcommand{\ee}{\end{eqnarray}}
\newcommand{\bra}[1]{\left< #1 \right|}
\newcommand{\ket}[1]{\left| #1 \right>}

\begin{document}  
\title{Are All Boer-Mulders Functions Alike?}

\author{Matthias Burkardt and Brian Hannafious}
 \affiliation{Department of Physics, New Mexico State University,
Las Cruces, NM 88003-0001, U.S.A.}
\date{\today}

\begin{abstract}
Chirally odd generalized parton distributions (GPDs) 
and the Boer-Mulders
function provide valuable information about spin-orbit correlations
for quarks in nucleons and other hadrons. We compare results for
the relevant GPD $\bar{E}^q_T$ from a variety of
phenomenological models as well as recent lattice results. We
find that $\bar{E}^q_T>0$ for nucleons as well as  the pion and for
both $u$ and $d$ quark. As a result, the corresponding Boer Mulders 
functions are all expected to be negative.
The sign of $\bar{E}^q_T$ arises from the relative sign between the
upper and lower Dirac components for the quark wave functions.
\end{abstract}

\maketitle

\section{Introduction}
Recent experiments by the {\sc Hermes} collaboration have demonstrated
the existence of significant single-spin asymmetries (SSAs) in 
semi-inclusive deep-inelastic scattering (SIDIS) \cite{HERMES}.
The Sivers function \cite{sivers}
measured in these experiments indicates a
negative correlation between the transverse proton polarization
and the transverse momentum for ejected $u$ quarks, while
the same correlation for $d$ quarks turned out to be positive
(Note that we adhere to the Trento convention \cite{Trento}).

Meanwhile, significant advances have been made in the theoretical
understanding of SSAs \cite{ansel}. 
For example, simple model calculations have 
illustrated very clearly that SSAs do not need to vanish in the
Bjorken limit \cite{BHS}. However, it has also become clear that
several ingredients need to conspire in order to produce SSAs
\cite{fengji}:
essentially what is needed is an interference between final state
interaction phases for quarks originating from different partial 
waves in the target wave function. Clearly, such a complex interplay
between different aspects of hadron structure not only makes
SSAs interesting, but it also presents a challange for theorists
to produce predictions that do not depend sensitively on model
parameters. Indeed, due to the complexity of these caclulations,
it even took a while until agreement was 
reached in observables as simple as the sign of the Sivers function
in the scalar diquark model.

Significant clarification regarding the sign issue was accomplished 
once it was noticed that the sign of these SSAs can be intuitively 
related to the sign of the deformations of certain spin-dependent
impact parameter dependent parton distributions \cite{IJMPA,sign}: 
the final
state interactions (FSIs) are expected to be attractive, on average, and
therefore impact parameter dependant parton distributions (IPDs), 
which are purely intrinsic
properties of hadrons, can be related to SSAs, which also invole
FSIs.  In the case of the Sivers effect, the relevant IPD is the Fourier transform of the Generalized
Parton Distribution (GPD) $E^q(x,0,-{\bf \Delta}_\perp^2)$ 
that describes the $x$-decomposition of the Pauli form factor 
$F^q_2(-{\bf \Delta}_\perp^2)$. Since the latter is approximately
known for $q=u,d$, it was
possible to predict the sign of the $u$ and $d$ quark Sivers 
functions \cite{mb1,mb2,mb3,metz} which were 
subsequently confirmed by the {\sc Hermes} data \cite{HERMES}.
 
Even when the target nucleon is unpolarized, the momentum 
distribution of its quarks ejected in SIDIS
can already exhibit a left-right 
asymmetry of the transverse quark momentum ${\bf k}_T$
relative to their own transverse spin ${\bf S_q}$ \cite{BM}
\bea
f_{q^\uparrow/p}(x,{\bf k}_T) = \frac{1}{2}\left[f_1^q(x,k_T^2)
-h_{1}^{\perp q}(x,k_T^2) \frac{ ({\bf {{\hat P}}}
\times {\bf k}_T)\cdot { {\bf S_q}}}{M}\right] .
\label{eq:BM}
\eea
${\bf {{\hat P}}}$ is a unit vector opposite to the direction of the
virtual photon momentum and $M$ is the target mass. 
This asymmetry can be measured experimentally by tagging the
transverse polarization of quarks produced in SIDIS using the
Collins effect \cite{collins}. Alternatively, one can also study 
the $\cos{2\phi}$ 
asymmetry in the unpolarized Drell-Yan process,
where $\phi$ is the azimutal angle of the $\mu^+\mu^-$ plane
about the virtual photon axis w.r.t. the incident proton.
Like the Sivers function,
the ``Boer-Mulders function'' $h_{1}^{\perp q}(x,k_T^2)$
 requires a combination of
orbital angular momentum with a difference between the final 
state interaction from different parts of the initial
quark orbit. Since the Boer-Mulders function specifies
the correlation between the transverse momentum asymmetry
of the struck quark and its spin, it provides important information
about the correlation between quark orbital angular momentum
and spin. Similar to the case of the Sivers function, one
expects that the sign of $h_1^{\perp q}$ can be related to the
distribution of transversely polarized quarks in impact parameter
space \cite{mb:BM}, which can be obtained from the Fourier transform
of chirally odd GPDs
\begin{eqnarray}
	\int \frac{dz^-}{4\pi} e^{ixP^+z^-}
	\bra{p^\prime}\bar{q}(-\frac{1}{2}z)\sigma^{+j}\gamma_5q(\frac{1}{2})\ket{p} 
&=&
	\frac{1}{2P^+} \biggl [ 
		H_T(x,\xi,t)\bar{u}\sigma^{+j}\gamma_5u + \widetilde{H}_T(x,\xi,t)		
\bar{u}\frac{\epsilon^{+j\alpha\beta}\Delta_\alpha P_\beta}{M^2}u \nonumber \\
		&& + E_T(x,\xi,t)\bar{u}\frac{\epsilon^{+j\alpha\beta}\Delta_\alpha\gamma_\beta}{2M}u
		+ \widetilde{E}_T(x,\xi,t)\bar{u}\frac{\epsilon^{+j\alpha\beta P_\alpha\gamma_\beta}}{M}u
	\biggr ] \\
&=&	\frac{1}{2P^+} \biggl [
		\left( H_T - \frac{t}{2M^2}\widetilde{H}_T\right)\bar{u}\sigma^{+j}\gamma_5u
		+ \widetilde{H}_T \bar{u}
\frac{\Delta^j\sigma^{+\alpha}\gamma_5\Delta_\alpha -
			\Delta\sigma^{j\alpha}\gamma_5\Delta_\alpha}{2m^2}u \nonumber \\
		&& + (E_T + 2\widetilde{H}_T)\bar{u}\frac{\epsilon^{+j\alpha\beta}\Delta_\alpha\gamma_\beta}{2M}u
		+ \widetilde{E}_T(x,\xi,t)\bar{u}\frac{\epsilon^{+j\alpha\beta P_\alpha\gamma_\beta}}{M}u
	\biggr ] \label{F_T_def}.
\end{eqnarray}
The linear combination $\bar{E}_T = E_T + 2\widetilde{H}_T$ multiplying the last term in (\ref{F_T_def})
describes the transverse deformation of the distribution of transversely polarized quarks in an
unpolarized nucleon \cite{DH}.  In ref \cite{mb:BM} it has been 
suggested that $h_1^{\perp q}$ and $\bar{E}_T^q$
have opposite signs.  Therefore, knowledge of $\bar{E}_T^q$ make 
qualitative predictions for
the Boer-Mulders function possible. However, not even the sign of $\bar{E}_T^q$ is
known experimentally. The main purpose of this paper is to develop some intuition about
the expected sign for $\bar{E}_T$ by analyzing this observable in a
variety of hadron models.

\section{Impact Parameter Dependent PDFs in the Bag model}
The well known MIT bag model offers a familiar venue to begin our discussion of the sign of SSAs
in the nucleon.  Bag model wavefunctions have the form
\begin{equation}
\Psi_m = 
\frac{N}{\sqrt{4\pi}}\left(
 \begin{matrix}
  ij_0 \chi_m \\
  -j_1 (\vec{\sigma} \cdot \hat{x}) \chi_m
 \end{matrix}
\right),
\end{equation}
where $\chi_m$ is a Pauli spinor, and the $j_n$ are the 
spherical Bessel functions and ${ N}$ is a normalization
factor.   In order to
slightly broaden the scope of our discussion, note that $j_1$ related to $j_0$ by a derivative. In fact, in the bag model, this relation
between the upper and lower component is a consequence of 
the (free) Dirac equation. With this in mind, let us instead focus
on a more general Dirac spinor,
\begin{equation}
\label{spinor}
\Psi_m = 
\left(
 \begin{matrix}
  if \chi_m \\
  -g (\vec{\sigma} \cdot \hat{x}) \chi_m
 \end{matrix}
\right),
\end{equation}
where $f$ is a monotonically decreasing radial function, 
and $g$ is the derivative of $f$.  

The impact parameter dependent parton distributions that we would
like to evaluate are of the form
\be
F_\Gamma (x,{\bf b}_\perp ) =
{\cal N}^{-1} \int \frac{dz^-}{4\pi} e^{ixp^+z^-} 
\left\langle p^+ , {\bf 0}_\perp \right| \bar{q}(0,{\bf b}_\perp)
\Gamma {q}(z^-,{\bf b}_\perp)\left| p^+, {\bf 0}_\perp
\right\rangle.
\ee
Here $\left| p^+, {\bf 0}_\perp \right\rangle$ is an eigenstate
of light-cone momentum $p^+$ as well as of the transverse
center of momentum operator ${\bf R}_\perp$ (for a definition of the latter see Refs. \cite{soper,IJMPA}).

From the bag model wave functions one can easily evaluate position
space densities of quark bilinears, but evaluating parton 
distributions requires introducing additional recipies in order to
deal with light-like correlation functions. This issue can be 
entirely avoided by focusing on the lowest moment of GPDs, 
yielding a (local) density that only depends on ${\bf r}_\perp$.
\be
\int dx F_\Gamma (x,{\bf b}_\perp) = (2p^+{\cal N})^{-1}
\left\langle p^+ , {\bf 0}_\perp \right| \bar{q}(0,{\bf b}_\perp)
\Gamma {q}(0,{\bf b}_\perp)\left| p^+, {\bf 0}_\perp
\right\rangle.
\label{local}
\ee
Instead of evaluating the matrix element of the operator in Eq. 
(\ref{local})  at the  origin between plane wave states, one may
equivalently localize also the longitudinal position of the state 
and instead integrate the operator over longitudinal position, 
yielding
\be
\int dx F_\Gamma (x,{\bf b}_\perp) = const. \int dx^3
 \left\langle  {\bf {\vec 0}}\right| \bar{q}(x^3,{\bf b}_\perp)
\Gamma {q}(x^3,{\bf b}_\perp)\left| {\bf {\vec 0}}
\right\rangle,
\ee
where $\left| {\bf {\vec 0}} \right\rangle$ denotes a nucleon 
state localozed at the origin (in all three space directions). 
Note that, as the bag model is not
boost invariant, there is a certain arbitrariness in the extraction
of $\bar{E}_T$ from bag model wave functions. In particular, the
reference point for the impact parameter ${\bf b}_\perp$, which
should be the transverse center of longitudinal momentum, is
here taken to be simply the center of the spherical bag.
However, since we are
only concerned about the overall sign of $\bar{E}_T$, such issues 
should not be significant.

Quarks with transverse polarization ${\bf s}$ are projected
out by the operator $\frac{1}{2}\bar{q}\left[\gamma^+ 
-s^ji\sigma^{+j}\gamma_5\right]q$ \cite{DH} and therefore the
vector field representing the transverse quark polarization density
is given by $-i\bar{q}\sigma^{+j}\gamma_5 q$. We thus consider
impact parameter dependent PDFs with $\Gamma = -i \sigma^{+j}\gamma_5$
which are related to the (Fourier transforms of the) chirally odd
GPDs $\bar{E}_T$, $H_T$ and $\tilde{H}_T$ \cite{DH} 
\be
F_T^i= -\varepsilon^{ij}b^j\frac{1}{M}{\bar{\cal E}}_T' 
+ S^i\left({\cal H}_T - \frac{1}{4M^2}\Delta_b \tilde{\cal H}_T
\right) + \left(2b^ib^j-b^2\delta^{ij}\right) S^j
\frac{1}{M^2}\tilde{\cal H}_T''
\ee
where we used script letters to denote the Fourier transforms of 
these GPDs and $S^j$ is the spin of the target. Only the term 
involving $\bar{\cal E}_T$ contributes
for an unpolarized target (averaging over transverse target
polarization), which is why it is only the GPD
$\bar{E}_T$ that is expected to be
(qualitatively) related to the BM function.
In the bag model, we extract this term by considering the
density corresponding to $\Gamma=-i\sigma^{+j}\gamma_5$ and
summing over the target spin. For a single quark state this 
procedure yields
\begin{eqnarray}
 \sum_m \bra{PS_{m}}\bar{\Psi}(x^3,\mathbf{b}_\perp)
	i\sigma^{+i}\gamma_5\Psi(x^3,\mathbf{b}_\perp) \ket{PS_{m}} 
 &=& \label{bm_exp} -\frac{1}{\sqrt{2}} \sum_m
	(f^2 + g^2) s^i_m + 2fg\epsilon^{ij}\hat{b}_\perp^j - 2g^2\hat{b}_\perp^i
		(\hat{b}_\perp\cdot \vec{s}_m) 
\label{eq:fg}
\end{eqnarray}
where $\vec{s}_m$ is the spin vector corresponding to the pauli 
spinor $\chi_m$.  The first and last terms of (\ref{bm_exp})
do not survive the sum over `target' polarizations. The asymmetry is given entirely by the middle term,
which is an interference between the uppper and lower components of 
(\ref{spinor}).  For the lowest moment of $\bar{\cal E}_T,$ we find
\begin{equation}
\kappa_T=
\int\! dx \bar{E}_T(x,0,0) =\int\! dx d^2{\bf b}_\perp 
\bar{\cal E}_T
	=
	\frac{2MG}{3\sqrt{2}\pi} \int_0^{R_0} dr \  r^3 fg.
	\label{res}
\end{equation}
The right hand side of (\ref{res}) is always positive because $f$ and $g$
are non-negative functions for $r$ less than the bag radius, implying that $\bar{\cal E}_T\geq 0$.

In the bag model,
the correlation between quark spin and quark 
orbital motion is the same, regardless of the orientation of $j_z$.
All quark spin orientations thus contribute coherently to 
$\bar{\cal E}_T^q$ and in the case of $d$ quarks, 
$\bar{\cal E}_T^d$ is 
equal to $\bar{\cal E}_T^d$ for a single quark, while for $u$ quarks
it is twice as large. In fact, for any model where the quarks
are confined by some mean field potential one finds that all
quark orbits give the same contribution to 
$\bar{\cal E}_T^q$ and thus $\bar{\cal E}_T^q$ is equal to 
$\bar{\cal E}_T^q$ for a single quark orbit, multiplied by the
number of quarks of flavor $q$. In particular, in the large $N_C$ limit, where
$N_u=N_d+1\rightarrow \infty$, the lowest $x$ moment of
$\bar{\cal E}_T^q$ is the same for $u$ and $d$ quark and both are of
order ${\cal O}(N_C)$. Since the support of GPDs shrinks to
$x = {\cal O}(1/N_C)$, this implies that
$\bar{E}_T^u(x,\xi,t)=\bar{E}_T^d(x,\xi,t)= {\cal O}(N_C^2)$

\begin{figure}
\unitlength1.cm
\begin{picture}(6,8)(3,2)
 	\includegraphics{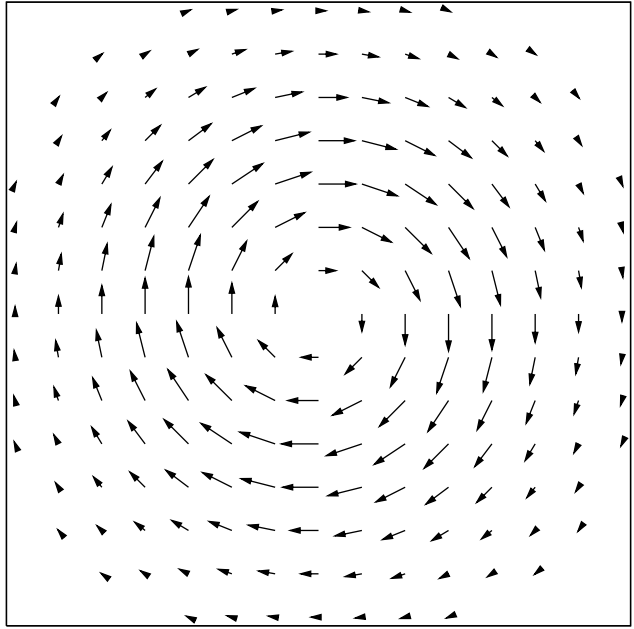}
\end{picture}
	\caption{Lowest moment of the impact parameter dependent transversity distribution
for an unpolarized target in the MIT bag model 
(\ref{eq:bag_vec_field}). The `outside' of the
spherical bag corresponds to the regions without arrows.  }
	\label{bag_vec_field}
\end{figure}

In order to visualise the transverse spin - impact parameter
correlation in the bag model, the vector field 
\be 
-\int dx^3 fg\epsilon^{ij}\hat{b}^j
\label{eq:bag_vec_field}
\ee
representing the lowest moment of the transversity density in an
unpolarized target has been plotted in Fig. \ref{bag_vec_field} 
for bag model wave functions $f=j_0(r)$, and
$g=j_1(r)$.
In the bag model, we thus optain a counter-clockwise polarization
for impact parameter dependent quark distributions. Upon invoking
the ``chromodynamic lensing'' mechanism (attractive FSI), 
and using the Trento convention \cite{Trento}, this
implies a negative Boer-Mulders function $h_1^\perp$. 
This result is also consistent with a direct calculation
of $h_1^\perp$ for the bag model in Ref. \cite{feng}.
The most important
question here is: how general is this result regarding the sign
of $h_1^\perp$, i.e. how much does it hinge on specific features of 
the bag model.

In order to address this issue, let us consider the general case of
a particle confined
by a combination of scalar potential (i.e. a mass term $m(r)$ that
depends on the radius) as well as a (zero component of a) 
vector potential $V(r)$. The
bag model corresponds to the limiting
case where the vector potential vanishes and the scalar potential
has the shape of an infinite square well.
For a free quark, the lower component $\phi_l$ is related to the
upper component $\phi_u$, via the free Dirac equation
\be
\phi_l = \frac{1}{E+m} {\vec \sigma}\cdot {\vec p} \phi_u.
\ee
In the presence of an external potential this relation changes
only slightly
\be
\phi_l = \frac{1}{E+m(r)-V(r)} {\vec \sigma}\cdot {\vec p} \phi_u.
\ee
The vector part of the confinement potential cannot exceed
the scalar part --- otherwise one encounters the Klein paradox. 
Therefore, $E+m-V$ should in general be positive 
As a result the sign in this relation is unchanged by the 
presence of the potentials $m(r)$ and $V(r)$. In particular, since
one would expect that the upper component $f(r)$ for the ground state
is a monotonically falling positive function
(its overall phase has been taken positive here), one expects $g(r)$
to be negative, regardless of the details of the potentials.
Note that while $E+m-V$ can be negative at isolated points without 
giving rise
to the Klein paradox, e.g. very close
to the origin for an electron moving in a Coulomb potential,
this does not invalidate our main point regarding the average
value of the product $f(r)\cdot g(r)$ or $f(r)\cdot g(r)r$ relevant
for the lowest moments of $\bar{E}_T$ at zero momentum transfer.

This discussion illustrates that the sign of the spin-orbit 
correlation (counterclockwise)
described by Eq. (\ref{eq:fg}) should be the same for the 
ground state
of all confining potential models.

\section{Diquark Models }
One of the main disadvantages of the bag model is its lack of
boost invariance, which results in certain ambiguities in
the parton interpretation of bag model wave functions.
In diquark models the distribution of quarks in a `nucleon'
is obtained by allowing it to split into a quark and
a spectator `diquark'. 
As the wave function is usually calculated perturbatively 
using a point-like vertex, boost invariance and electromagnetic 
gauge invariance are straightforward to maintain. The same applies
to (perturbatively calculated) QCD final state interactions between
the quark and the diquark.
Because of the latter features, it is thus not surprising that
the scalar diquark model has provided the first clear and convincing 
example for the fact that SSAs can survive in the Bjorken limit 
\cite{BHS}.

The phenomenological motivation for diquark models is the idea that,
as long as the momentum transfer on the spectators is not very large,
one may replace them by a single, point-like degree of freedom ---
the diquark. The quark-quark interaction in the nucleon is more 
attractive in the scalar channel with quark spins anti-alligned.
This further motivates the assumption that the spectator diquark
carries scalar quantum numbers. In this work, we will mostly focus
on the scalar diquark model only, although more sophisticated
diquark models exist.

The scalar diquark model wave function, has only two spin components
corresponding to positive ($\uparrow$) and negative quark helicity
($\downarrow$). Angular 
momentum conservation implies that the wave function components for
`nucleon' states with
positive ($\Uparrow$) 
helicity take on the form
\bea
\psi^\Uparrow_\uparrow &=& f(x,{\bf k}_\perp^2) \\
\psi^\Uparrow_\downarrow &=& (k^x+ik^y) 
g(x,{\bf k}_\perp^2) \nonumber
\eea
Time reversal implies for the negative helicity states
($\Downarrow$) 
\bea
\psi^\Downarrow_\uparrow(x,{\bf k}_\perp) &=& (-k^x+ik^y) g(x,{\bf k}_\perp^2)\\
\psi^\Downarrow_\downarrow(x,{\bf k}_\perp) &=& f(x,{\bf k}_\perp^2)\nonumber .
\eea
For transversely polarized quarks (lower index)
in a transversely polarized target (upper index)
this implies
\bea
\psi^{+\hat{x}}_{+\hat{x}}(x,{\bf k}_\perp) &=& f(x,{\bf k}_\perp^2) + 
ik^y g(x,{\bf k}_\perp^2) \\
\psi^{+\hat{x}}_{-\hat{x}}(x,{\bf k}_\perp) &=& -
k^x g(x,{\bf k}_\perp^2),
\eea
i.e. in impact parameter space
\bea
\label{eq:diq}
\tilde{\psi}^{+\hat{x}}_{+\hat{x}} (x,{\bf c}_\perp)
&\equiv&
\int d^2{\bf k}_\perp e^{i{\bf k}_\perp \cdot {\bf c}_\perp}
\psi^{+\hat{x}}_{+\hat{x}} (x,{\bf k}_\perp)=
\tilde{f}(x,{\bf c}_\perp^2) + \frac{\partial}{\partial c^y}
\tilde{g}(x,{\bf c}_\perp^2)
\\
\tilde{\psi}^{+\hat{x}}_{-\hat{x}} (x,{\bf c}_\perp)
&\equiv&
\int d^2{\bf k}_\perp e^{i{\bf k}_\perp \cdot {\bf c}_\perp}
\psi^{+\hat{x}}_{-\hat{x}} (x,{\bf k}_\perp) = 
i\frac{\partial}{\partial c^x}
\tilde{g}(x,{\bf c}_\perp^2)
\nonumber
\eea
As a result of the interference between the
helicity non-flip $\tilde{f}$ and
helicity-flip component $\partial_y\tilde{g}$, 
the distribution of quarks with the
same transverse polarization as the `nucleon' 
$\left|\tilde{\psi}^{+\hat{x}}_{\hat{x}} (x,{\bf c}_\perp)\right|^2$
exhibits a transverse
asymmetry proportional to $\tilde{f}\partial_y\tilde{g}$. 
The wave function component where the transverse quark spin
is opposite to the nucleon spin does not exhibit such an asymmetry.
Intuitively, one can understand the origin of the asymmetry in the
scalar diquark model as follows.
When the `nucleon' splits into a quark and a diquark, the orbital
angular momentum of the two particle system can be either $l=0$ or
$l=1$ (Note that this is not in conflict with parity
invariance as upper and lower components of quark spinors transform
differently under parity and leading twist quark densities 
result from superpositions of upper and lower components).
Higher orbital angular momenta are not possible as the total
angular momentum (OAM plus spin of the quark) must still be
$\frac{1}{2}$. For quarks that have the same (transverse) 
spin as the `nucleon',
the orbital wave function can have either $l=0$ or $l=1$ and 
$l_x=0$. The asymmetry results from the interference between these
two wave function components very much in the same way as
hybridization in chemistry results in asymmetric electron orbitals.
In contradistinction, the wave function component 
where the quark spin is opposite to the `nucleon' spin
is pure $l=1$, i.e. there cannot be any interference between $l=0$
and  $l=1$ and hence no asymmetry.
This very simple observation also `explains' why the BM and Sivers 
functions are identical in the scalar diquark model
(for an explicit calculation of both functions in this model
see Ref. \cite{boer}): a left-right 
asymmetry only occurs in the wave function component where the 
transverse spin of the quark and the transverse spin of the `nucleon'
are the same. Thus the correlation between the transverse position
space asymmetry with the quark spin is the same that it is with the
nucleon spin. Since (at least in diquark models) the transverse
momentum space asymmetry in SIDIS arises entirely from the initial
transverse position space asymmetry, the BM function (representing
the correlation between transverse momentum and spin of the struck
quark) and the Sivers function (representing
the correlation between transverse quark momentum and  `nucleon'
spin) must be the same. While it has been known that those two
T-odd functions are the same in the scalar diquark model, the
intuitive explanation provided here is new.

For models containing axial vector diquarks the situation is mode 
complicated, as there can also be interference between wave
function components where the transverse quark spin is
opposite to the `nucleon' spin (and the vector diquark spin is
along the `nucleon' spin direction). As a result, the correlation
between quark spin and its transverse position is no longer 
identical to the correlation between `nucleon' spin and quark
transverse position. Hence the Sivers and BM functions in vector
diquark models do not have to be identical.

In the scalar diquark model
\bea
\psi^{+\hat{y}}_{+\hat{y}} (x,{\bf c}_\perp) &=&
\left[M+\frac{m}{x}+ \frac{1}{x}
\frac{d}{dc^y} \right] \phi({\bf c}_\perp^2) 
\nonumber\\
\psi^{+\hat{y}}_{-\hat{y}} (x,{\bf c}_\perp) &=& -\frac{i}{x}
\frac{d}{dc^y} \phi({\bf c}_\perp^2) 
\eea
where ${\bf c}_\perp$ is the distance between the active
quark and the spectator and $\phi({\bf c}_\perp)$ is monotonically 
decreasing with ${\bf c}_\perp^2$. This result is of the above
form (\ref{eq:diq}), 
with $\tilde{f} = \left[M+\frac{m}{x} \right]\phi$ and
$\tilde{g}= \frac{1}{x}\phi$, i.e. quarks polarized in the
$+\hat{x}$ direction are shifted towards negative $y$. Again,
we find a counter-clockwise pattern for the transverse quark 
polarization corresponding to a positive GPD $\bar{E}_T^q$.
A qualitatively similar result is obtained for axial vactor diquark models.

It is easy to understand why we obtain the same sign as in the bag 
model. The quark interacts with the diquark (to form the nucleon)
only when the two are at the same point, i.e. in impact parameter
space the interaction is a contact interaction. For
nonzero impact parameter, the quark is thus a free particle and as
such subject to the free Dirac equation relating
upper and lower component. Just as was the case in the bag model,
this determines the relevant phase which in turn determines the
sign of $\bar{E}_T^q$. This observation has far reaching consequences
for other (e.g. axial vector) diquark models. What all diquark models
have in common is the fact that quark and diquark interact only
when they merge to become the nucleon, i.e. the quark-diquark
interaction is local. Both for scalar and axial vector diquarks,
the the quark-diquark state is mainly an $s$ state and therefore
the same general statements regarding the sign of $\bar{E}_T^q$
apply to both the axial vector and scalar case.

\section{Relativistic Constituent Quark Models}
In relativistic constituent quark models for hadron structure,
one starts from nonrelativistic forms for the quark wave function.
Nontrivial spin structure is then generated by boosting these
wave functions to the infinite momentum frame. In this procedure,
using non-interacting boost operator, the net result is that
instant form  spin eigenstates 
$\left| k^3, {\bf k}_\perp; \pm \right\rangle_I$
are replaced by light-front helicity eigenstates 
$\left| k^+, {\bf k}_\perp; \pm \right\rangle_F$
using what is often referred to as a Melosh transformation
\cite{melosh}
\bea
\left| k^3, {\bf k}_\perp; + \right\rangle_I 
&=& w \left[ (k^+ +m)
\left| k^+, {\bf k}_\perp; + \right\rangle_F
-k^R
\left| k^+, {\bf k}_\perp; - \right\rangle_F\right]
\nonumber\\
\left| k^3, {\bf k}_\perp; - \right\rangle_I 
&=& w \left[ (k^+ +m)
\left| k^+, {\bf k}_\perp; - \right\rangle_F
+k^L
\left| k^+, {\bf k}_\perp; + \right\rangle_F\right]
\eea
where $w=1/\sqrt{2k^+(m+k^0)}$, 
$k^{R/L} = k^1 \pm ik^2$, and $k^+ = k^0+k^3$.
For quarks with nonzero transverse momentum, the boost to the
infinite momentum frame (or equivalently the Melosh transformation)
naturally introduces a corelation between quark spin and quark 
orbital angular momentum and a non-trivial spin structure of the
nucleon state --- even if the original (instant form) wave function
only contained s-wave components and SU(6) wave functions.

Such models have been used  to estimate $\bar{E}_T^q$ in Ref. 
\cite{pasquini}. Studying a wide range of model parameters, it
was found that $\kappa^u_\perp$ in the range between $1.98$ and
$3.60$, and $\kappa^d_\perp$ in the range between $1.17$ and $2.36$.
The sign of these results is in agreement with the results 
from the other models. Furthermore, similar to the bag model, one 
finds $\kappa_u \approx 2\kappa_d$. It is possible to 
understand the origin of the sign for $\bar{E}_T$ in this model.
The relativistic constituent models start from pure $s$ wave 
wavefunctions. Any $p$ wave component is generated by
the Melsoh transformation --- which employs the free (noninteracting)
boost operator. It is thus not surprising that the phase relation
between the $s$ and $p$ wave is thus the same as for all the other
models studied here.

\section{$\bar{E}_T$ in the Pion}
Since the pion has no spin, its impact parameter dependent PDFs
are much simpler than in the case of the nucleon. The distribution
of quarks with spin $s^i$ in impact parameter space reads

\begin{equation}
\frac{1}{2}\left[ F + s^i F_T^i \right] = 
 H(x,\mathbf{b}^2) + s^i \epsilon^{ij} b^j \frac{2}{m} \frac{\partial}{\partial \mathbf{b}^2} \bar{E}_T(x,\mathbf{b}^2),
 \label{E_pion}
\end{equation}
where ${\cal H}(x,\mathbf{b}^2)$ and $\bar{\cal E}_T(x,\mathbf{b}^2)$
are again the fourier transforms of the GPDs $H(x,0,t)$ and
$\bar{E}_T(x,0,t)$ respectively whose definition 
is particularly simple 
\begin{eqnarray}
& & \int \frac{dz^-}{4\pi} e^{ixP^+z^-} \bra{\pi^\prime} \bar{q}(-\frac{1}{2}z) \gamma^+
   q(\frac{1}{2}z) q\ket{\pi} \vert_{z^+=0,z=0} 
= H(x, \xi, t) \\
& & \int \frac{dz^-}{4\pi} e^{ixP^+z^-} \bra{\pi^\prime} \bar{q}(-\frac{1}{2}z)
   \sigma^{+j} \gamma_5 q(\frac{1}{2}z)\ket{\pi} \vert_{z^+=0,z=0} 
= \frac{1}{\Lambda} \bar{E}_T(x,\xi, t) \frac{\epsilon^{+j\alpha\beta}\Delta_\alpha P_\beta}{P^+}.
\end{eqnarray}
Here $\Lambda$ is some hadronic mass scale, 
which needs to be included in the definition if one wants 
$\bar{E}_T(x,\xi, t)$ to be dimensionless (cf. Eq. \ref{F_T_def}). 
For the nucleon, the
natural choice is the nucleon mass, but chosing $\Lambda=m_\pi$
for the pion would perhaps not be a wise choice, 
as this would unnecessarily 
complicate the discussion of the chiral limit. Since we are again 
mainly interested in the sign of $\bar{E}_T$, 
we leave this issue
open, but perhaps $\Lambda =4\pi f_\pi$ or $\Lambda=M_N$ would
be more useful choices for the pion case.

Except for a slight change in the bag radius, the quark wave 
functions in the bag model are the same for pions and nucleons. 
Therefore, apart from a slight rescaling due to the different
bag radii, $\bar{E}^u_T$ in a $\pi^+$ is the same as
$\frac{1}{2}\bar{E}^u_T$ or $\bar{E}^d_T$ in the proton.
Here the factor $\frac{1}{2}$ accounts for the fact that there
are twice as many $u$ quarks in a proton than in a $\pi^+$.

As an alternative model for GPDs in the pion we also considered
the Nambu-Jona-Lasino (NJL) model. In this model, the
quark and the anti-quark interact via a $\gamma_5$ contact 
interaction in the $s$-channel, and form factors or GPDs are
obtained by evaluating momentum intergrals containing 
quark traces of the form
\be
\frac{
tr\left[\gamma_5 (k+\frac{\Delta}{2}+M)\Gamma
(k-\frac{\Delta}{2}+M)\gamma_5 (k-p+M)\right]}
{\left[\left(k+\frac{\Delta}{2}\right)^2-M^2\right] 
\left[(k-p)^2-M^2\right]
\left[\left(k-\frac{\Delta}{2}\right)^2-M^2\right]
} .
\ee
Here $M$ is the (constituent) quark mass
and $\Gamma$ depends on the current under consideration.
For example, $\Gamma=\gamma^+$ is used to calculate
PDFs and the GPD $H$, yielding for the trace
\be
tr[\Gamma=\gamma^+]= 4\left\{ p^+(M^2-k^2) +k^+[M^2- (p-k)^2+p^2\right].
\ee
Evaluating the expression when the spectator is on-shell
[$(p-k)^2=M^2$], which arises from complex countour intagration 
for $k^-$, and dropping all terms proportional to $m_\pi^2$ this
yields 
\be
tr[\Gamma=\gamma^+]= 4p^+\frac{M^2+{\bf k}_\perp^2}{p^+-k^+}
= 4\frac{M^2+{\bf k}_\perp^2}{1-x}.
\ee
Together with a factor $\frac{1}{1-x}$ in the energy denominators,
this gives rise to the well known result of a constant PDF in the 
NJL model (for $m_\pi=0$).

In the calculation of $\Delta_i \bar{E}_T$, it is
$\Gamma=\gamma^+\gamma^i$ which enters the above trace.
For simplicity, we focus here on $\bar{E}_T$ for $\Delta\rightarrow
0$ which determines the average spin dipole moment $\kappa_\perp$.
Since the denominator is even in $\Delta$, only the 
$\Delta$-dependence in the numerator matters in this limit, and
we thus consider only the relevant numerator
\be
tr[\Gamma=\gamma^+\gamma^i] &=&
tr\left[(k+\frac{\Delta}{2}+M)\gamma^+\gamma^i
(k-\frac{\Delta}{2}+M)(p-k+M)\right]
= \frac{\Delta_i}{2} tr\left[[(k+M)\gamma^+ + \gamma^+(k+M)]
(p-k+M)\right]\nonumber\\
&=& 4\Delta_i M p^+,
\ee
yielding again $\bar{E}_T>0$. A more detailed analysis yields
$\bar{E}_T\propto (1-x)$, for $m_\pi\rightarrow 0$, where the 
constant of proportionality depends on the cutoff in the loop 
integrals. 

Enlightened by our discussion about the sign of $\bar{E}_T$ in the
nucleon, it is easy to uderstand the result in the NJL model:
the interaction between quark and antiquark in this model is
also a contact interaction, i.e. the quark and antiquark
follow the free Dirac equation, except when they sit on top of each
other. Therefore, by the same reasoning as in the bag model,
$\bar{E}_T$ should be positive, as the explicit calculation 
confirmed. Lattice calculations yield the same sign \cite{latticepi}.

\section{Summary and Conclusions}
We have studied the chirally odd GPD $\bar{E}_T$ for both $u$ and $d$
quarks in a variety of 
models (Bag model, potential models,
diquark model, NJL-model, relativistic constituent quark models).
All models provide the same sign $\bar{E}_T>0$.
This sign also agrees with recent results from lattice QCD 
calculations. While the physical origin of the sign is obscure in
lattice QCD calculations, it is much more transparent in the model
calculations. As the quark helicity flips in the matrix elements
contributing to $\bar{E}_T$, the nucleon helicity does not. Hence
the quark orbital angular momentum between the initial and final
state must differ by one unit and therefore $\bar{E}_T$ arises from
the interference between wave function components that differ by
one unit of orbital angular momentum. In many models for the 
structure of ground state hadrons only $s$ and $p$ wave components 
are present and higher OAM is either completely absent or negligible.
Moreover, in these models, the $p$-wave component arises 
primarily from the `lower' component of the hadron wave function,
which is related to the upper component via the Dirac equation,
which thus determines the relative phase between the upper and 
lower components. 
While the presence of interactions tends to modify the relation
between the upper and lower components quantitatively, it does not 
change the phase relation and all models yield the same phase 
relation between upper and lower component that also holds for
free quarks. Lattice results confirm this result. Of course, one
needs to be cautious as present day lattice calculations are
still performed at moderately high quark masses. The pion masses
used in Ref. \cite{latticeETbar} range from $m_\pi^2 \approx
1\,\mbox{GeV}^2$ to $m_\pi^2 \approx 0.2\,\mbox{GeV}^2$.
and the fact that the phase relation between upper and lower
components is the same as for the free Dirac equation could in
principle be an artifact of the large quark mass. However, the
results from Ref. \cite{latticeETbar} are also rather stable
against variations of the quark mass, and the
sign of $\bar{E}_T^q$ does not change upon extrapolation to the
physical quark masses, yielding $\kappa_T^u\approx 2.93$ and
$\kappa_T^u\approx 1.90$

Such a unanimous agreement between such a variety of theoretical
approaches to an observable that has never been measured is
probably almost unprecedented. While some of the model
predictions for the sign of $\bar{E}_T^q$ existed before, what
is new in this work is the intuitive understanding about the sign 
of $\bar{E}_T^q$
as resulting from the relation between upper and lower components
in the (free) Dirac equation. Despite all the complexities of QCD, 
lattice calculations yield results that are qualitatively consistent 
with quarks in an $s$ state that have a $p$-wave in the lower 
(Dirac) component of the wave function and with a relative phase
obtained from the free Dirac equation. While this agreement may be a 
pure accident, it is still very tempting to interpret this
result as an indication that a nontrivial fraction of the orbital 
angular momentum in the nucleon wave function is merely resulting
from the $p$-wave admixture in the lower component of quark 
spinors in a bound state.

{\bf Acknowledgements:}
This work has been partially supported by the DOE under grant number 
DE-FG03-95ER40965. 

\bibliography{diquark3.bbl}
\end{document}